\documentclass[conference]{IEEEtran}
\IEEEoverridecommandlockouts
\usepackage{cite}
\usepackage{amsmath,amssymb,amsfonts}
\usepackage{amsthm}

\usepackage{algorithm}
\usepackage{algpseudocode}
\usepackage{graphicx}
\usepackage{textcomp}
\usepackage{xcolor}
\usepackage{physics}
\usepackage{mathtools}
\usepackage{tabularx}
\usepackage[export]{adjustbox}
\usepackage[capitalise]{cleveref}
\usepackage{tcolorbox}
\tcbuselibrary{theorems}

\newtcbtheorem[]{mytheo}{Theorem}%
{colback=green!5,colframe=green!35!black,fonttitle=\bfseries}{th}
\newtcbtheorem[]{mycorollary}{Corollary}%
{colback=brown!5,colframe=brown!35!black,fonttitle=\bfseries}{th}

\def\BibTeX{{\rm B\kern-.05em{\sc i\kern-.025em b}\kern-.08em
    T\kern-.1667em\lower.7ex\hbox{E}\kern-.125emX}}
\begin{document}

\title{Analysis and Improvement of Rank-Ordered Mean Algorithm in Single-Photon LiDAR
\thanks{The work is supported, in part, by the DARPA / SRC CogniSense JUMP 2.0 Center, NSF IIS-2133032, and NSF ECCS-2030570. William C. Yau completed the research in summer 2024 at Purude University through the Summer Undergraduate Research Fellowship (SURF).}
}
\author{\IEEEauthorblockN{William C. Yau\textsuperscript{1}, Weijian Zhang\textsuperscript{2}, Hashan Kavinga Weerasooriya\textsuperscript{2}, Stanley H. Chan\textsuperscript{2}}
\IEEEauthorblockA{\textsuperscript{1}\textit{Department of Physics and Department of Computer Sciences, UC Berkeley, Berkeley CA, U.S.A.}  \\
\textsuperscript{2}\textit{School of Electrical and Computer Engineering, Purdue University, West Lafayette IN, U.S.A.} \\
\textsuperscript{1}{yauwilliam69@berkeley.edu}, \textsuperscript{2}\{zhan5056, hweeraso, stanchan\}@purdue.edu}
}

\maketitle

\begin{abstract}
Depth estimation using a single-photon LiDAR is often solved by a matched filter. It is, however, error-prone in the presence of background noise. A commonly used technique to reject background noise is the rank-ordered mean (ROM) filter previously reported by Shin \textit{et al.} (2015). ROM rejects noisy photon arrival timestamps by selecting only a small range of them around the median statistics within its local neighborhood. Despite the promising performance of ROM, its theoretical performance limit is unknown. In this paper, we theoretically characterize the ROM performance by showing that ROM fails when the reflectivity drops below a threshold predetermined by the depth and signal-to-background ratio, and its accuracy undergoes a phase transition at the cutoff. Based on our theory, we propose an improved signal extraction technique by selecting tight timestamp clusters. Experimental results show that the proposed algorithm improves depth estimation performance over ROM by 3 orders of magnitude at the same signal intensities, and achieves high image fidelity at noise levels as high as 17 times that of signal.
\end{abstract}


\section{Introduction}
Active optical methods, such as Light Detection and Ranging (LiDAR) systems, are rapidly advancing thanks to the advent of Single-photon Avalanche Diode (SPAD) sensors, achieving tens-of-picosecond time resolution \cite{b9}. A typical LiDAR system utilizes a periodically pulsed laser to measure distances between the system and the scene. For each pixel, over a period called the \textit{dwell time}, a LiDAR system collects timestamps of photon detections relative to the most recent pulse emission and constructs a histogram. The time delay is proportional to the \textit{depth} of the scene. The amplitude of the histogram, proportional to the number of photons collected, is related to the \textit{reflectivity} \cite{b1}. Typically, accurate estimations require $10^2$ to $10^3$ photons per pixel (PPP) collected and the histogram to be binned finely \cite{b1}. However, such intensity is difficult to achieve when the scene is dark, the background noise is much stronger, or objects are rapidly moving.

Recent work has demonstrated how to exploit probabilistic models to form accurate depth and reflectivity images from an average PPP of 1.0 \cite{b1}. A key contribution of \cite{b1} is the rank-ordered mean (ROM) filter which exploits the underlying spatial continuity in natural scenes to remove noise. However, its notable limitations in darker pixels prompted our interest to study the theoretical limits of its performance. 

In additional to our theoretical findings, we propose a signal extraction method by leveraging the low-variance nature of signal photon arrivals. To further amplify signal, we aim to employ pixel neighborhood data borrowing and duplication. We aim to make accurate imaging possible when signal intensity is as low as 0.2 PPP, and noise intensity as high as 17 times that of signal. This improvement in photon-efficiency and noise-tolerance is amenable to parallelization and avoids building a histogram like \cite{b2}.

\begin{figure}[t]
\centerline{\includegraphics[width=1.02\linewidth, left]{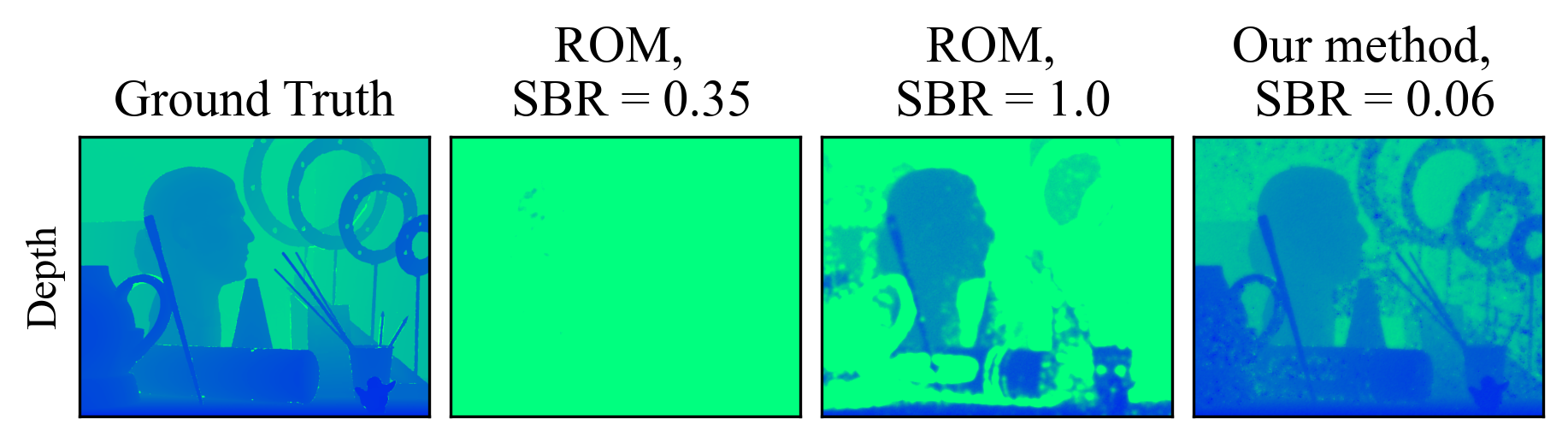}}
\caption{Comparison between our proposed method in \Cref{section:difference} and the ROM filter \cite{b1} under different signal-to-background ratios (SBR).}
\label{fig:teaser}
\end{figure}

\subsection*{Main Contributions}
\begin{enumerate}
    \item \textit{Theoretical Analysis of ROM Filtering:} We theoretically derive the conditions under which ROM estimates are accurate. We accurately predict a phase transition from failure to success, with an estimate on absolute error.
    \item \textit{Neighborhood Consensus Filtering:} We propose an improvement of ROM based on a local consensus of the minimal time differences. We show that this method is more noise-tolerant and photon-efficient than existing methods.
\end{enumerate}

\section{Data Acquisition, and Probabilistic Modeling}
The goal of 3D image reconstruction is to form a reflectivity image $\boldsymbol{\alpha} \in \mathbb{R}_{\geq 0}^{N_i \times N_j}$ and depth image $\boldsymbol{z} \in \mathbb{R}_+^{N_i \times N_j}$ of the scene. Each pixel is represented by indices $(i, j)$, $i \in \{1, \dots, N_i\}$, $j \in \{1, \dots, N_j\}$. The depth associated with a pixel $(i, j)$ is the distance between the imager and the scene $z_{i, j} \in [0, z_{\max})$, where $z_{\max}$ is the maximum detectable depth subject to hardware specifications. The reflectivity $\alpha_{i, j} \in [0, 1]$ is the ratio between the irradiance received at pixel $(i, j)$ compared to that of signal. This paper mostly focuses on depth reconstruction. 

\subsection{Illumination and Detection}
We assume the LiDAR system emits a periodically pulsed laser light with period $T_r$ towards the scene. For each pixel, in each period, a photon-flux waveform described by $s(t)$, $t \in [0, T_r)$ is incident upon the scene. To avoid distance aliasing, we assume $T_r > 2z_{\max}/c$, where $c$ is the speed of light. The root-mean-square (RMS) pulse width $T_p \ll 2z_{\max}/c$ is chosen for high depth resolution. Conventionally and in this paper, $s(t)$ is modeled as a Gaussian pulse with $\sigma = \frac{T_p}{2}$. Imaging accuracy increases when the number of pulses $N \in \mathbb{N}_0$ increases. The total time required for a measurement is the \textit{acquisition period} $T_a = NT_r$.

\subsection{Probabilistic Model of Back-reflected Photons}
After being illuminated by pulse $s(t)$, pixel $(i,j)$ reflects photon flux towards the detector. The sensor generates time-resolved single-photon detections called \textit{clicks}. The sensor's \textit{quantum efficiency} $\eta \in [0, 1)$ controls the ratio of photons registered compared to light flux received. Clicks are represented in timestamps $t_{i, j}^{(l)}\in[0, T_r)$ denoting the detection instant relative to the latest fired pulse, where $l$ is an index assigned at registration. 

The arrival of photons is modeled as a Poisson process, and the rate of photon detections in a repetition period is
\begin{align}
    \lambda_{i, j}(t) = \eta \alpha_{i, j} s(t - t_{i, j}^*) + (\eta b_{\lambda} + d) \label{eq:lambda}
\end{align}
where $t_{i, j}^* = 2z_{i, j}/c$ denotes the expected signal arrival time for traveling back and forth a depth of $z_{i, j}$, $b_\lambda \in \mathbb{R}_{\geq 0}$ denotes the ambient light flux at the operating optical wavelength $\lambda$ of the imager, and $d \in \mathbb{R}_{\geq 0}$ denotes the detector dark count. 

Total photon count in a repetition period is defined as \(k_{i, j} \coloneq \int_0^{T_r} \lambda_{i, j}(t) \dd{t} = \eta \alpha_{i, j} S + B\), including background count \(B \coloneq (\eta b_{\lambda} + d)T_r\), and signal count $\eta \alpha_{i, j} S$ with \(S \coloneq \int_0^{T_r}s(t) \dd{t}\). Then, the scene-average signal-to-noise ratio $\text{SBR} \coloneq \eta \bar{\alpha} S/B$.

\subsection{Depth Estimation: Maximum Likelihood (ML) Estimation}
For each pixel $(i, j)$, assuming that a non-empty set of observed photon detection times $\mathcal{T}_{i, j} = \{t_{i, j}^{(l)}\}_{l = 1}^{k_{i, j}}$ is obtained, the constrained maximum likelihood (CML) estimator subject to the search space for depth $\boldsymbol{z}:z_{i, j} \in [0, z_{\max})$ is given as the log-matched filter \cite{b4}, 
\begin{equation}
    \hat{\boldsymbol{z}}^\text{CML} = \underset{\boldsymbol{z}}{\arg {\min}} \underbrace{\sum_{i=1}^{N_i} \sum_{j = 1}^{N_j} \sum_{l = 1}^{k_{i, j}} - \log \left [ s \left( t_{i, j}^{(l)} - 2z_{i, j}/c \right) \right ]}_{\coloneq \mathcal{L}(\boldsymbol{z})}
\end{equation} where $\mathcal{L}$ is the objective function.

 A more sophisticated approach takes into account of the assumption that an physical scene has to be a union of piecewise continuous and smooth blocks, giving rise to the penalized ML (PML) estimator \cite{b1}, where an additional regularization term $\text{pen}(\boldsymbol{z})$ has to be minimized along with $\mathcal{L}({\boldsymbol{z}})$:
\begin{align}
    \hat{\boldsymbol{z}}^{\text{PML}} = \underset{\boldsymbol{z}}{\arg \min} \, \mathcal{L}(\boldsymbol{z}) + \beta \, \text{pen}(\boldsymbol{z}) 
    \label{eq:PML}
\end{align}
The regularizer $\text{pen}(\boldsymbol{z})$ has to be convex and penalizes non-smoothness in $\boldsymbol{z}$. The weight $\beta > 0$ controls the severeness of penalization. The total variation semi-norm is a popular choice in image reconstruction for its edge-preserving property \cite{b5}.

\begin{figure}[t]
\includegraphics[width=1.0\linewidth, left]{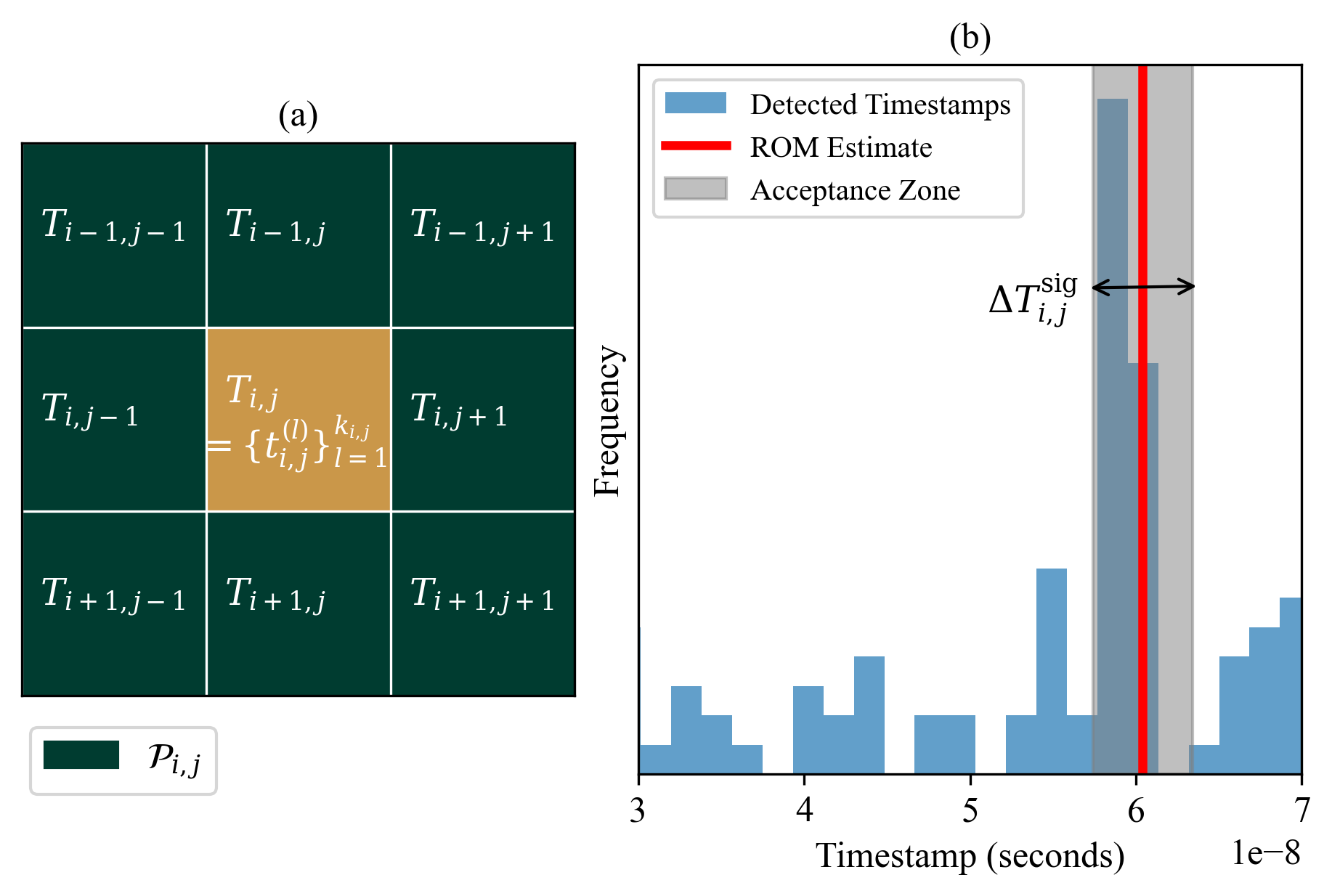}
\caption{(a): How ROM works: $8$ neighboring pixels (dark green) of target pixel $(i, j)$ (center, orange) form $\mathcal{P}_{i, j}$ and carry independent timestamps that are combined to form $\mathcal{T}_{i, j}$. Timestamps belonging to $(i, j)$ are left out. (b): Timestamp censoring of ROM, where only timestamps within $\Delta T_{i, j}^{\text{sig}}$ centered at the ROM estimate $t_{i, j}^{\text{ROM}}$ are taken into account for depth estimation (Figure 2(b) is not drawn in scale).}
\label{fig:ROM_explain}
\end{figure}

\subsection{Denoising: The ROM Filter}
The PML estimator suffers from bias and high-variances when the signal-to-background ratio is low. Various approaches have been reported to tackle the problem of signal-background separation, i.e. to extract signal detections from noisy data. The extracted data are then fed to the PML estimator for higher reconstruction accuracy. The rank-ordered mean (ROM) filter stands out as a popular choice \cite{b1}. 

ROM combines timestamps of $8$ neighboring pixels for each pixel $(i, j)$ and forms a grand set of detections $\mathcal{T}_{i, j} = \bigcup_{(p, q) \in \mathcal{P}_{i, j}} T_{p, q}$, where $\mathcal{P}_{i, j}$ denotes the $3 \times 3$ neighborhood of $(i, j)$, as illustrated in \Cref{fig:ROM_explain}. The median timestamp among $\mathcal{T}_{i, j}$ is chosen to be the ROM estimate $t_{i, j}^{\text{ROM}}$. 
Subsequently, extract the set of presumed signal detections by choosing timestamps close enough to $t_{i, j}^{\text{ROM}}$:
\begin{equation}
    T^\text{sig}_{i, j} = \Big \{ t \in T_{i, j}: |t - t_{i, j}^{\text{ROM}}| < \frac{1}{2} \underbrace{\frac{4 T_p B}{\eta \hat{\alpha}_{i, j}S + B}}_{\coloneq \Delta T_{i, j}^{\text{sig}}} \Big \} \label{eq:censor}
\end{equation}
where $\Delta T_{i, j}^{\text{sig}}$ is the zone of acceptance centered at $t_{i, j}^{\text{ROM}}$ that increases when the background detection probability increases. This is to accommodate for less reliable ROM estimates that happen when the reflectivity estimate $\hat{\alpha}_{i, j}$ at that pixel is lower \cite{b1}.

\section{Analysis of the ROM Filter}
Although the ROM filter improves depth estimations for scenes with high SBRs, it fails on low reflectivity pixels as noted in \cite{b1}. In the theorem below, we propose a necessary condition under which ROM works:

\begin{figure}[t]
    \centerline{\includegraphics[width=1.0\linewidth, right]{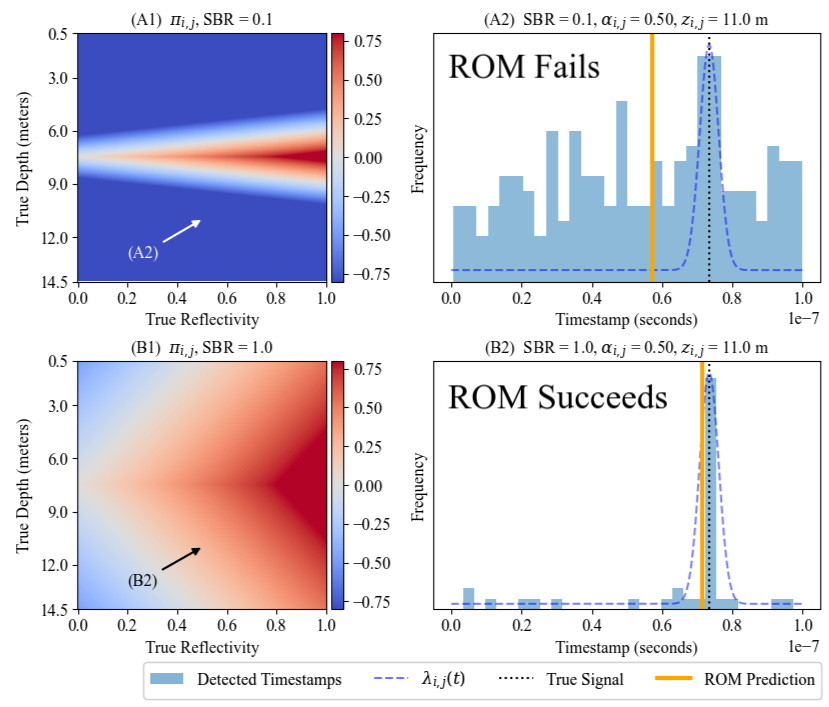}}
    \caption{Behavior of ROM estimate in two typical scenarios. Figure (A1) and (B1) displays predictor $\pi_{i, j}$ for each pixel for the toy scene with SBR = 0.1 and 1.0 respectively. Histograms (A2) and (B2) shows timestamps collected for the same pixel under each case. (A2) denotes a failure of ROM as increased background count pulls ROM estimator towards the halfway time $t_{\frac{1}{2}} = cT_r/4$, while (B2) denotes a success.}
    \label{fig:annotated_histograms}
\end{figure}

\begin{mytheo}{}{threshold}
Under the assumptions outlined in \Cref{tab:assumptions}, for pixel $(i, j)$ with true reflectivity $\alpha_{i, j}$ and true depth $z_{i, j}$, the ROM estimate $t_{i, j}^{\text{ROM}}$ equals the true signal timestamp $t^* = \frac{2z_{i, j}}{c}$ up to a precision of $\frac{1}{2}T_p \ll T_r$ only when:
\begin{equation}
     \frac{\alpha_{i, j}}{\bar{\alpha}/\text{SBR}} \geq   \frac{|z_{i, j}-z_{\frac{1}{2}}|}{z_{\frac{1}{2}}} \label{eq:theorem}
\end{equation}
where $z_{\frac{1}{2}} = cT_r/4$ is the halfway depth, $\bar{\alpha}$ is the scene-average reflectivity, and SBR is the scene-average signal-to-background ratio.
\end{mytheo}

Intuitively, ROM works by assuming the signal count is large enough to skew the 50-th timestamp $t_{i, j}^{\text{ROM}}$ towards the true signal timestamp $t_{i, j}^* = 2z_{i, j}/c$. However, signal count $S \propto \alpha_{i, j}$, and its ability to skew the median relies on its ratio over the background count $B \propto \bar{\alpha}/\text{SBR}$. Effectively, the left hand side of \eqref{eq:theorem} denotes a per-pixel signal-to-background ratio, and can be interpreted as the "strength" of the signal. A low-reflectivity pixel produces a weaker signal amplitude relative to background and tends to fail.

The right hand side denotes a "hurdle" the signal has to overcome. Naturally, even without signal, the median lies at the halfway time $t_{\frac{1}{2}} = T_r/2$ of time range $[0, T_r)$, which corresponds to halfway depth $z_{\frac{1}{2}} = \frac{T_r}{2} \frac{c}{2} = \frac{cT_r}{4}$. The degree of skew required by the median scales with the difference between $t_{\frac{1}{2}}$ and $t^*_{i,j}$. The further $t^*_{i,j}$ is from $t_{\frac{1}{2}}$, the stronger the signal is required, hence the numerator $|z_{i, j}-z_{\frac{1}{2}}|$. A special case happens at $z_{i, j} = z_{\frac{1}{2}}$ where the ROM is accurate for any $\alpha_{i, j}$ and SBR, since the median naturally lies at $t_{\frac{1}{2}}$.

Our goal is to investigate the necessary conditions for $t_{i, j}^{\text{ROM}} \approx t_{i, j}^*$. Since \eqref{eq:censor} allows timestamps that are sufficiently close to $t_{i, j}^{\text{ROM}}$ to survive, we want to investigate conditions for $t_{i, j}^{\text{ROM}}$ to be within an interval of width $T_p$ centered at $t^*_{i, j}$.

ROM assumes that for pixel $(i, j)$, reflectivity and depth for pixels in $\mathcal{P}_{i, j}$ are approximately uniform, i.e. true depth and reflectivity images $\boldsymbol{z}$, $\boldsymbol{\alpha}$ are piecewise constant. Hereby we assume they all share $\alpha = \alpha_{i, j}$, $z = z_{i, j}$ and drop the subscript $(i, j)$ for notational simplicity. 

Since signal pulse width $T_p \ll T_r$, $s(t) \sim \mathcal{N}(t^*, T_p)$ sharply peaks inside $[t^* - \frac{1}{2}T_p, t^* + \frac{1}{2}T_p]$ and quickly dies off to $0$ elsewhere in $[0, T_r)$. Other assumptions include a constant and uniform rate of background detections $B/T_r$ over the entire scene and over time, low photon flux ($\leq 1$ per pulse) such that dead time is negligible \cite{b7, b8}, and a unique depth in $[0, cT_r/2)$ for each pixel. \Cref{tab:assumptions} summarizes the assumptions for this analysis and algorithms introduced in \Cref{section:improvements}.

\begin{table}[t]
  \centering
  \caption{Assumptions of Scene, Background \& Signal Properties}
  \label{tab:assumptions}
  \begin{tabularx}{\linewidth}{X}
  \hline \hline
  Assumptions\\ [0.5ex] 
  \hline
  1. A unique depth in $[0, cT_r/2)$ for each pixel \\ 
  2. Background rate $B/T_r$ is uniform over the entire scene \\
  3. $B$ is constant and uniform over time \\
  4. $B = \eta b_{\lambda} + d$ is known from calibration \\ 
  5. Low photon flux ($\leq 1$ per pulse) throughout the scene \\
  6. Signal pulse width $T_p \ll T_r$ \\
  7. True depth and reflectivity images $\boldsymbol{z}$, $\boldsymbol{\alpha}$ are piecewise constant \\
  \hline \hline
  \end{tabularx}
\end{table}

\begin{proof}
Since ROM chooses the median among timestamps collected, it is helpful to analyze the photon count in regions separated by $t^*$.
Recall that \eqref{eq:lambda} models the rate of photon arrival within $[0, T_r)$. Total photon count in $[0, t^* - \frac{1}{2}T_p]$ is
\begin{align}
     k_{-} = \int_0^{t^*- \frac{T_p}{2}} \lambda(\tau) \dd{\tau} &= \eta \alpha \int_0^{t^* - \frac{T_p}{2}} s \left(\tau -t^*  \right) \dd{\tau} + B\frac{t^*}{T_r} \nonumber \\
     &= B\frac{t^* - T_p/2}{T_r}. \label{eq:lamb_LHS_rslt}
\end{align}
because $s(t) =0$ outside $[t^* - \frac{1}{2}T_p, t^* + \frac{1}{2}T_p]$. 

Similarly, photon count in $[0, t^* + \frac{1}{2}T_p]$ is
\begin{align}
     k_{+} &= \eta \alpha \int_{0}^{t^* +\frac{T_p}{2}}  \left[  s \left(\tau -t^*  \right) + \frac{B}{T_r} \right] \dd{\tau} \nonumber \\ 
     & = \eta \alpha S + B\frac{t^*+T_p/2}{T_r}. \label{eq:lamb_RHS_rslt}
\end{align}
because $t^* \in [0, t^* + \frac{1}{2}T_p]$ so signal count $\eta \alpha S$ is included.

Although there are $N$ pulses in an acquisition period, $\lambda(t)$ remains constant. $t^{\text{ROM}}$ is independent of $N$ as the median is constant with scalar multiplication.

Recall that total photon count $k = \eta \alpha S + B$. $t^{\text{ROM}}$ is the median among them, meaning there are $k/2$ detections in both ranges $[0, t^{\text{ROM}})$ and $[t^{\text{ROM}}, T_r)$. For $t^{\text{ROM}} \approx t^*$, this bisection condition has to be necessarily fulfilled alongside with conditions \eqref{eq:lamb_LHS_rslt} and \eqref{eq:lamb_RHS_rslt}, i.e., 
\begin{equation}
 B\frac{t^* - T_p/2}{T_r} \leq \frac{1}{2} (\eta \alpha S + B) \leq \eta \alpha S + B\frac{t^*+T_p/2}{T_r}. \label{eq:sandwich}
\end{equation}Terms with $T_p$ can be dropped as $T_p \ll T_r$.
Given a scenario with scene-average reflectivity $\bar{\alpha}$, and a constant scene-average $\text{SBR} = \eta \bar{\alpha} S/B$, theorem \ref{th:threshold} directly follows from \eqref{eq:sandwich}. 
\end{proof}
For convenience, we define predictor 
\begin{equation}
    \pi_{i, j} \coloneq \frac{\alpha_{i, j}}{\bar{\alpha}/\text{SBR}} - \frac{|z_{i, j}-z_{\frac{1}{2}}|}{z_{\frac{1}{2}}}  
\end{equation}
which indicates that the ROM estimate is accurate when $\pi_{i, j} \geq 0$, but fails when $\pi_{i, j} < 0$. 

\begin{mycorollary}{}{corollary}
The absolute error of ROM estimates are
\begin{equation}
    \left |t^{\text{ROM}} - t^* \right |_{i, j} = \max \left (- \frac{T_r}{2}\pi_{i, j}, 0 \right )
\end{equation}
\end{mycorollary}

\begin{proof}
    If $t^{\text{ROM}} \geq t^*+T_p/2$, i.e. $z_{\frac{1}{2}} > z$:
\begin{align}
    t^{\text{ROM}} - t^* = \frac{\frac{k}{2} - k_+}{B/T_r} = - \frac{T_r}{2} \left( \frac{\alpha_{i, j}}{\bar{\alpha}/\text{SBR}} + \frac{z - z_{\frac{1}{2}}}{z_{\frac{1}{2}}} \right)
\end{align}
A similar expression can be obtained for $t^{\text{ROM}} \leq t^*-T_p/2$, i.e. $z_{\frac{1}{2}} < z$. Combining, we have
\begin{align*}
    \left |t^{\text{ROM}} - t^* \right |_{i, j} = - \frac{T_r}{2} \left( \frac{\alpha_{i, j}}{\bar{\alpha}/\text{SBR}} - \frac{|z_{i, j} - z_{\frac{1}{2}}|}{z_{\frac{1}{2}}} \right) = - \frac{T_r}{2}\pi_{i, j} \label{eq:absolute_error}
\end{align*}
However, the ROM estimate is accurate for $\pi_{i, j} \geq 0$. For all $\pi_{i, j}$, $\left |t^{\text{ROM}} - t^* \right |_{i, j} = \max \left (- \frac{T_r}{2}\pi_{i, j}, 0 \right)$.
\end{proof}

\section{Improvement of the ROM Filter} \label{section:improvements}
In broad terms, ROM fails when the signal is not strong enough compared to background, or central enough for the median. We present the following methods to improve ROM.

\subsection{The Mode Filter} \label{section:mode}
Since ROM fails when the centrality requirement is not met, a natural solution is to consider the mode instead of the median. When the signal amplitude is strong, yet too different from $cT_r/4$, directly recognizing the timestamp count peak gives a better estimate of where the signal is (assuming that signal intensity is stronger than the background). Timestamps are continuous variables, so binning is required for deciding the most populous timestamp. A precision of $\sim T_p/2$ proves to be sufficiently accurate. The subsequent censor step is the same as \eqref{eq:censor} with $t_{i, j}^{\text{mode}}$ replacing $t_{i, j}^{\text{ROM}}$. 

\begin{algorithm}[t]
\caption{The Neighborhood Consensus Filter}\label{alg:difference}
\begin{algorithmic}
\footnotesize
\Procedure{NeighborhoodFormation}{}
\State Round up $\frac{16}{\eta S \bar{\alpha}N}$ to nearest square of an odd integer $\widehat{N}_{\text{sp}}$
\State Define side length $n_{\text{sp}} = \sqrt{\widehat{N}_{\text{sp}}}$
\For {pixel $(i, j)$}
    \State Form neighborhood $\mathcal{N}_{i, j}$ as square of side length $n_{\text{sp}}$, center $(i, j)$
    \State \textbf{\textit{Store}} grand set of timestamps $\mathcal{T}^{\text{sp}}_{i, j} = \{ t^{(u)}_{i, j}\}_{u=1}^{\tilde{k}_{i, j}}$ as the union of $\{ t^{(l)}_{x, y}\}_{l=1}^{k_{x, y}}$ for all $(x, y) \in \mathcal{N}_{i, j}$
\EndFor
\EndProcedure
\Procedure{TimestampDifferences}{$\mathcal{T}^{\text{sp}}= \{ t^{(u)}\}_{u=1}^{\tilde{k}}$}
    \State Sort $\{ t^{(u)}\}_{u=1}^{\tilde{k}}$ in ascending order to form $(t^{(u)})_{u=1}^{\tilde{k}}$
    \State Form $(d^{(u)})_{u=1}^{\tilde{k} - 1}$ with $d^{(u)} \gets t^{(u + 1)} - t^{(u)}$
    \State Form $(c^{(u)})_{u=1}^{\tilde{k} - 3}$ with $c^{(u)} \gets \frac{1}{4}d^{(u)} + \frac{1}{2}d^{(u+1)} +  \frac{1}{4}d^{(u+2)}$
    \If{$\min c^{(u)} \geq T_p$} $T^\text{sig} \gets \emptyset$ \Comment{Ensure signal selected}
    \Else
        \State $u_{\min} \gets {\arg \min}_{u} \, c^{(u)}$
        \State $u_{\min} \gets u_{\min} +2$ \Comment{Index realignment}
        \State $t^{\text{diff}} \gets t^{(u_{\min})}$
        \For {$t \in \mathcal{T}^{\text{sp}}$}
            \If {$|t - t^{\text{diff}}| < T_p$} Append $t$ to $T^\text{sig}$
            \EndIf
        \EndFor 
        \State \textbf{\textit{Store}} $T^\text{sig}$
    \EndIf
\EndProcedure
\For {pixel $(i, j)$} \Call{TimestampDifferences}{$\mathcal{T}^{\text{sp}}_{i, j}$} \EndFor
\Procedure{OutlierRejection}{}
\State Define $\mathcal{T}^\text{sig}$ as the union of all $T^\text{sig}_{i, j}$
\State Calculate $\bar{t}^{\text{sig}} \gets \text{mean of }\mathcal{T}^\text{sig}$, \, $\sigma_t^{\text{sig}} \gets \text{standard deviation of }\mathcal{T}^\text{sig}$
\For {pixel $(i, j), t \in T^\text{sig}_{i, j}$}
    \If {$|t - \bar{t}^{\text{sig}}| \geq p\sigma_t^{\text{sig}}$} \textbf{\textit{Remove}} $t$ from $T^\text{sig}_{i, j}$
    \EndIf
\EndFor
\EndProcedure
\State $\hat{z}^{\text{PML}} \gets \text{PML}(\boldsymbol{T}^\text{sig})$ \Comment{PML Depth Estimation, eq. \eqref{eq:PML}}
\end{algorithmic}
\end{algorithm}

\subsection{The Neighborhood Consensus Filter}\label{section:difference}
It is computationally inefficient to conduct binning. We thus introduce the neighborhood consensus filter, which does not require such step. Inspired by \cite{b3}, we propose a simple neighborhood formation scheme beyond ROM's $3\times 3$ neighborhood to amplify signal intensity. We overcome the central tendency of ROM by selecting the most closely located signal clusters. Finally, an outlier rejection scheme is employed for clean-up. 

Data input to our algorithm are detection timestamps $\{t_{i, j}^{(l)}\}_{l = 1}^{k_{i, j}}$ for each pixel $(i, j)$. We assume knowledge of 1.) $B$, calibrated before measurement, and 2.) $S$, estimated by procedures well-known and outlined in \cite{b2}. Components of the algorithm is detailed below.

\begin{figure}[t]
    \centerline{\includegraphics[width=1.02\linewidth, right]{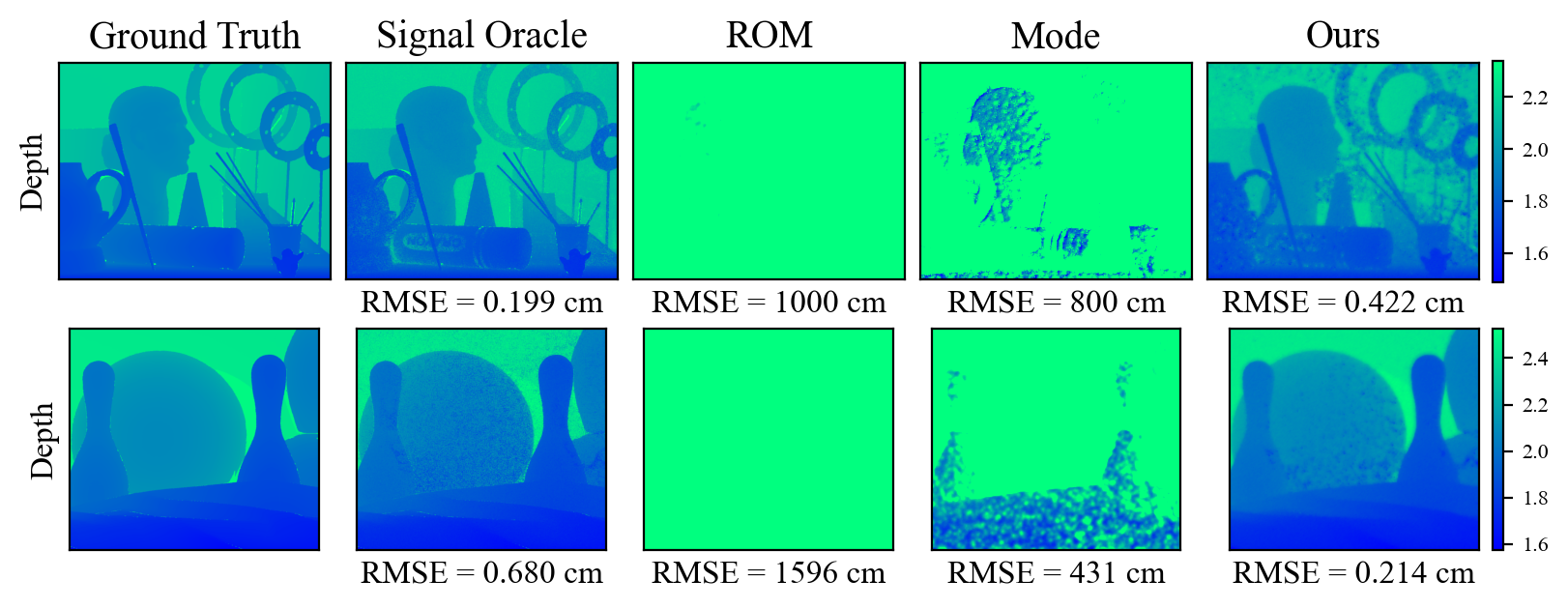}}
    \vspace*{-1mm}
    \caption{Simulated processing results for Art and Bowling scenes \cite{b6} at SBR = 0.2 and 2.0 signal PPP.}
    \label{fig:images}
    \vspace*{-5mm}
\end{figure}

\subsubsection{Neighborhood Formation} \label{section:neighborhood}
Based on empirical results, our method relies on $\geq 4$ signal count at each pixel, but in most cases the scene-average signal PPP $\sigma = \eta  \bar{\alpha} S N$ is below $4$. To compensate for this, we borrow timestamps from the neighborhood of $(i, j)$ with size $N_{\text{sp}}$ given by empirical equation \(N_{\text{sp}} \approx 16/\sigma\). 

To simplify procedures, we form a square neighborhood around $(i, j)$, which requires an odd-numbered side length. A simple way to find such side length is to round $N_{\text{sp}}$ up to the nearest perfect square of an odd integer. Let that be $\widehat{N}_{\text{sp}}$. Then the side length is $n_{\text{sp}} = \sqrt{\widehat{N}_{\text{sp}}}$. 

The neighborhood of pixel $(i, j)$, $\mathcal{N}_{i, j}$, is then defined as the square with side length $n_{\text{sp}}$ centered at $(i, j)$. We combine timestamps for each pixel $(x, y) \in \mathcal{N}_{i, j}$ to form a grand set of timestamps $\mathcal{T}^{\text{sp}}_{i, j} = \{ t^{(u)}_{i, j}\}_{u=1}^{\tilde{k}_{i, j}}$, where $u$ is the new index, and $\tilde{k}_{i, j} = \sum k_{x, y}$ is the neighborhood total photon count. These data are fed into the procedure outlined below.

\subsubsection{Timestamp Differences Calculation}
To detect signal timestamps, instead of using the median in ROM, or the mode in \Cref{section:mode}, we leverage differing characteristics of temporal distributions of background and signal photons. Background photons distribute uniformly in $[0, T_r)$ with variance $T_r^2/12$, but signal timestamps result from signal pulse $s(t-t^*)$, which is a Gaussian distribution strongly peaked at $t^*$ with variance $T_p^2/4 \ll T_r^2/12$. In other words, signal detections can only occur close to one another in a narrow time pocket. To leverage this fact, we look for closely located timestamp clusters by processing each $\mathcal{T}^{\text{sp}}_{i, j} = \{ t^{(u)}_{i, j}\}_{u=1}^{\tilde{k}_{i, j}}$. Below, the subscript $(i, j)$ is dropped for notational simplicity. 

Given $\{ t^{(u)}\}_{u=1}^{\tilde{k}}$, the promise of the clustering behavior of signal timestamps suggests checking the ones that are the closest together. To achieve this, we first sort them in ascending order to form an ordered sequence $(t^{(u)})_{u = 1}^{\tilde{k}}$ where $t^{(u)} \leq t^{(u+1)}$, and then calculate the consecutive differences $d^{(u)} = t^{(u + 1)} - t^{(u)}$, resulting in $(d^{(u)})_{u = 1}^{\tilde{k} - 1}$. 

Supposedly the minimum $d$ must belong to a pair of signal timestamps. However, a pair of background detections might incidentally come close together, defeating the signal pair to give the smallest $d$. To eliminate this possibility, we can build consensus from more neighboring timestamps. Signal timestamps, regardless of number, must all occur within a window of width $T_p$ centered at $t^*$, while it is increasingly rare for background timestamps forming increasingly populous clusters. Therefore, the strongest cluster must be the one with consistently low $d$ between more than two timestamps.

Therefore, consensus can be built by taking a moving weighted mean across $(d^{(u)})$, and selecting the minimum $d$. We choose to achieve this by applying a discrete linear convolution on $(d^{(u)})$ with the weight sequence $[\frac{1}{4}, \frac{1}{2}, \frac{1}{4}]$, resulting in a sequence $(c^{(u)})_{u= 1}^{k-3}$, where $c^{(u)} = \frac{1}{4}d^{(u)} + \frac{1}{2}d^{(u+1)} + \frac{1}{4}d^{(u+2)}$. We then select the order for smallest convoluted distance $u_{\min}={\arg {\min}}_{u} \, c^{(u)}$.

To know if a signal cluster has effectively been chosen, we check if $c^{(l_{\min})} \leq T_p$. If not, we report no estimates for this pixel. Otherwise, we choose signal estimate $t^{\text{diff}} = t^{(u_{\min} + 2)}$ because the indices have to realign to match with the original ones. We then extract signal timestamps by only choosing the ones that are less than $T_p$ from $t^{\text{diff}}$:
\begin{equation}
    \mathcal{T}^\text{sig} = \Big \{ t \in \{ t^{(u)}\}_{u=1}^{\tilde{k}}: |t - t^{\text{diff}}| < T_p \Big \} 
\end{equation}

\subsubsection{Outlier Rejection}
$t^{\text{diff}}$ can still be inaccurate when background count overwhelms the signal (i.e. low SBR), or signal count in low-reflectivity pixels is insufficient. If we directly provide $\mathcal{T}^\text{sig}$, possibly containing outliers, to PML depth estimation, the penalization term in \eqref{eq:PML}, instead of modifying the outliers to close in on the accurate values, might falsely change accurate values to adhere to the outliers,  contaminating the accurate estimates. 

To effectively block out these outliers, it is realistic to assume that the set of pixel true depths $\{z_{i, j}\}$ concentrates around a mean $\bar{z}$ with standard deviation $\sigma_z$. Hence, we calculate scene-average timestamp $\bar{t}^{\text{sig}}$ from all extracted timestamps $\{T^\text{sig}_{i, j}\}$ and standard deviation $\sigma_t^{\text{sig}}$, and reject timestamps with absolute error $|t - \bar{t}^{\text{sig}}| \geq p \sigma_t^{\text{sig}}$. $p$ is slightly scene-dependent, but a generic $p = 1$ is effective for low SBR cases. 

The entire neighborhood consensus filter is summarized in \Cref{alg:difference}.

\begin{figure}[t]
    \centerline{\includegraphics[width=\linewidth, left]{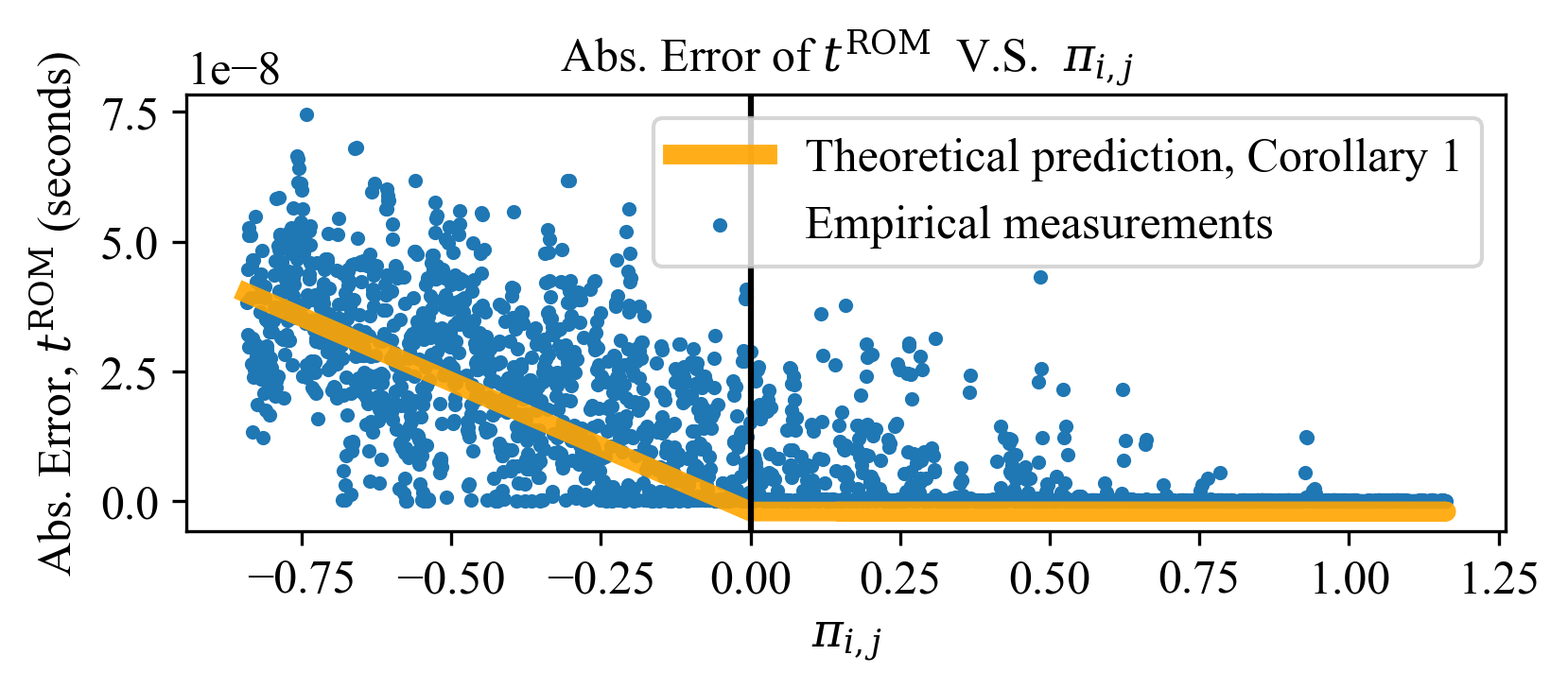}}
    \vspace*{-5mm}
    \caption{Absolute error of ROM estimates against predictor $\pi_{i, j}$. Timestamps were generated for the toy scene with SBR = 1.0 and 2.0 signal PPP.}
    \label{fig:error}
    \vspace*{-5mm}
\end{figure}

\section{Experiment}
We present experimental results based on simulated data to verify Theorem \ref{th:threshold} and demonstrate effectiveness of our proposed filter. Timestamps are simulated with the same parameters used in \cite{b3}. For the signal pulse, $T_p = 270$ ps, $T_r = 100$ ns, and $\eta = 0.35$. To simulate a low-signal scenario, $S = 0.0114$ such that on average one signal photon is generated with about $500$ pulses.

Signal counts were sampled from a Poisson distribution with mean $\eta \alpha_{i, j} S$. Signal detection times were simulated from a Gaussian distribution with mean $2z_{i, j}/c$ and $\sigma = T_p/2$. Background counts were sampled from a Poisson distribution with mean $B = \eta \bar{\alpha} S/\text{SBR}$, and their detection times are uniformly chosen over $[0, T_r)$. 

To quantify performance of a $N_i \times N_j$ estimate, we use the root-mean-square error $\text{RMSE}(\boldsymbol{z}, \hat{\boldsymbol{z}}) = \sqrt{||\boldsymbol{z}-\hat{\boldsymbol{z}}||_2^2/N_iN_j}$ in units of meters.

\subsection{Verifying Theorem \ref{th:threshold}}
Simulations were done on a $1000 \times 1000$ toy scene with linearly increasing reflectivity and depth along the horizontal and vertical axes respectively. Specifically, for pixel $(i, j), i \in \{1, \dots, 1000\}, j \in \{1, \dots, 1000\}$, the scene has true reflectivity $\alpha_{i, j} = j/1000$, and true depth $z_{i, j} = 0.5+i\frac{14}{1000}$ meters. We avoided nonphysical depths $0$ and $z_{\max} = 15$ meters.

We simulated timestamps with SBR = 1.0 and signal PPP = 2.0, and their ROM estimates were found. Their absolute errors from true signal timestamps $t_{i, j}^*$ were plotted against predictors $\pi_{i ,j}$ in \Cref{fig:error}, together with theoretically predicted errors using Corollary \ref{th:corollary}. We can clearly see the phase transition predicted in Theorem \ref{th:threshold}, and Corollary \ref{th:corollary} provides accurate error estimates to empirical data.

\begin{figure}[t]
    \centerline{\includegraphics[width=1.0\linewidth, right]{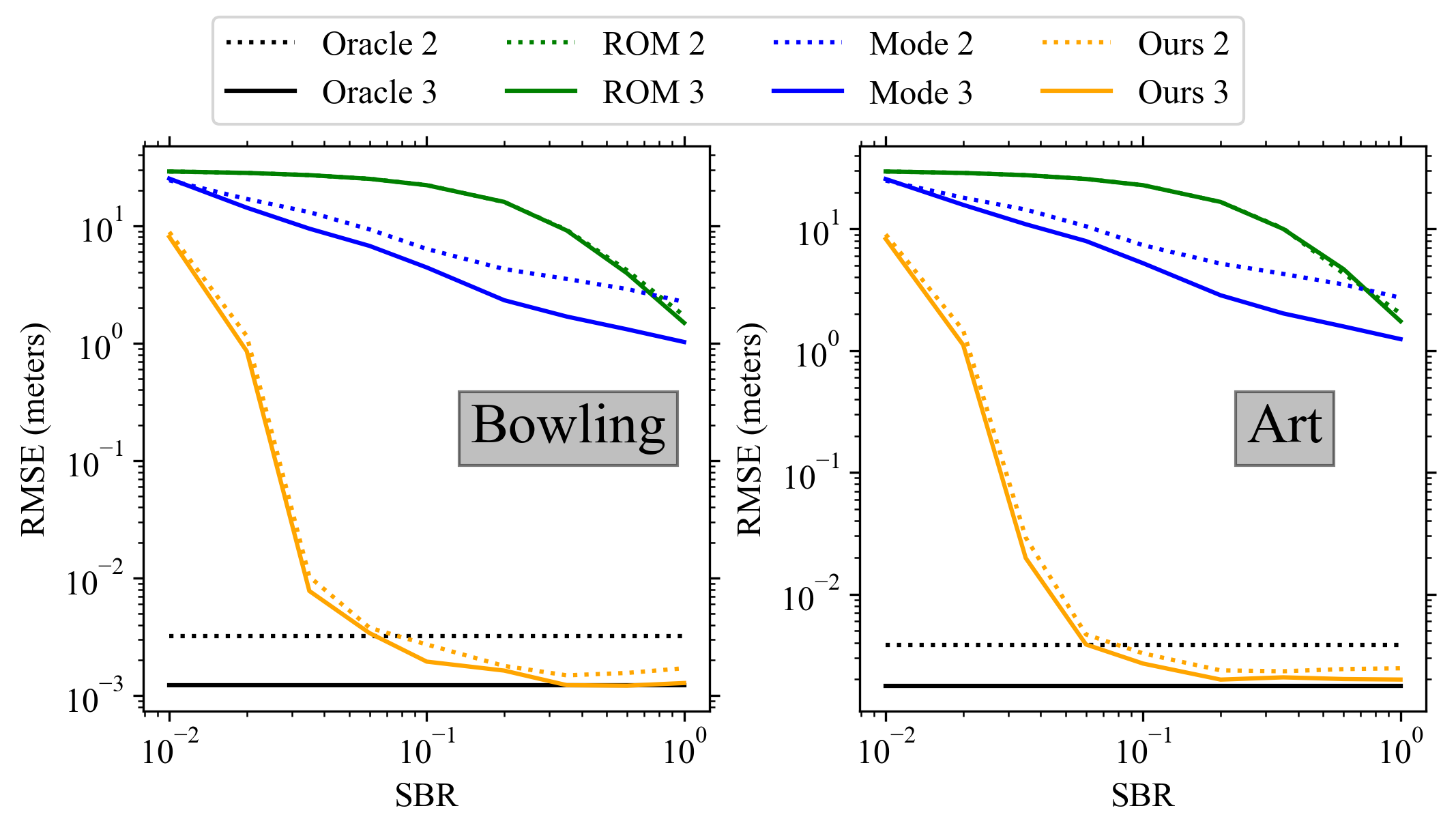}}
    \caption{Performance evaluation for depth estimations with signal PPPs of 2.0 and 3.0 and varying SBRs. Plotted are the average RMSEs of 10 trials.}
    \label{fig:plot_sbr}
    \vspace*{-5mm}
\end{figure}

\subsection{Comparing ROM, Mode, and Our Proposed Filter}
Datasets were simulated using physical scenes from the Middlebury dataset \cite{b6}. The Bowling and Art scenes were chosen respectively for simple and complicated scenes. ROM is compared with mode and our neighborhood consensus filter. An \textit{signal oracle} indicating perfect signal extraction are generated with timestamps simulated at SBR = $10^{12}$.

\Cref{fig:images} shows an exemplar result at SBR = 0.2 and 2.0 signal PPP. ROM images are blank as all pixels are out of bounds. The mode filter is able to extract some signals by directly selecting histogram peaks. Our proposed filter is the most effective as important object boundaries are preserved and depth estimates are accurate. 

\Cref{fig:plot_sbr} shows performance as a function of SBR. All methods improve with increasing SBR, but our method rapidly catches up with the oracle, displaying superior noise-tolerance.

\begin{figure}[t]
    \centerline{\includegraphics[width=1.0\linewidth, left]{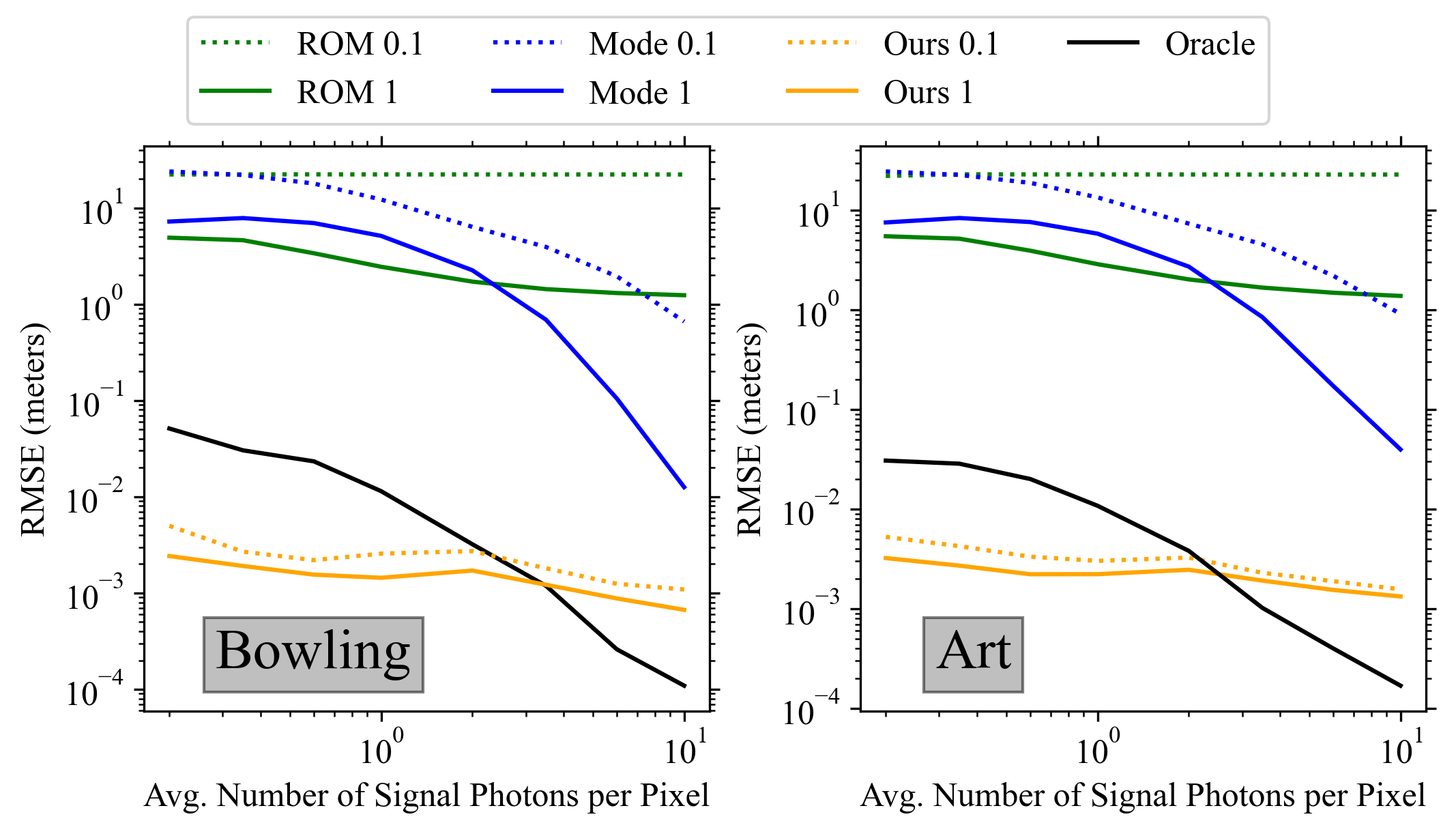}}
    \vspace*{-2.5mm}
    \caption{Performance evaluation for depth estimations with SBRs of 0.1 and 1.0 and varying signal PPPs. No neighborhood formation is done for Oracle, while our method has neighborhood size determined in \Cref{section:neighborhood}. Plotted are the average RMSEs of 10 trials.}
    \vspace*{-5mm}
    \label{fig:plot_sppp}
\end{figure}

\Cref{fig:plot_sppp} displays performance as a function of signal PPPs. Our method beats the signal oracle for low signal PPPs thanks to neighborhood formation, which duplicates accurately selected signal timestamps in same-depth neighborhoods, implying a superior signal efficiency. For the same signal PPP, our method is about 3 orders of magnitude better than ROM. The small fluctuations in \Cref{fig:plot_sbr,,fig:plot_sppp} are caused by discrete jumps in neighborhood sizes.

\section{Conclusion}
This paper proposed Theorem \ref{th:threshold} stating that ROM fails when reflectivity drops below a threshold dependent on depth and SBR. To overcome ROM's limitation, we proposed a new method that leverages the temporal closeness of signal timestamps, instead of only relying on signal count like ROM.  Additional steps increases accuracy by consensus-ensuring convolution and outliers rejection. Experimental results demonstrated successful reconstruction at SBR $\geq 0.06$ and increased photon efficiency of $\geq 3$ orders of magnitude. Future work includes improving reflectivity estimation and runtime analysis.


\vspace*{-1mm}
\begin{thebibliography}{00}
\bibitem{b1} D. Shin, A. Kirmani, V. K. Goyal and J. H. Shapiro, "Photon-Efficient Computational 3-D and Reflectivity Imaging With Single-Photon Detectors," in IEEE Transactions on Computational Imaging, vol. 1, no. 2, pp. 112-125, June 2015, doi: 10.1109/TCI.2015.2453093.
\bibitem{b2} A. Kirmani et al., “First-photon imaging,” Science, vol. 343, no. 6166,
pp. 58–61, Jan. 3, 2014.
\bibitem{b3} J. Rapp and V. K. Goyal, "A Few Photons Among Many: Unmixing Signal and Noise for Photon-Efficient Active Imaging," in IEEE Transactions on Computational Imaging, vol. 3, no. 3, pp. 445-459, Sept. 2017, doi: 10.1109/TCI.2017.2706028.
\bibitem{b4} D. L. Snyder, Random Point Processes. Hoboken, NJ, USA: Wiley, 1975.
\bibitem{b5} S. Osher, A. Solé, and L. Vese, “Image decomposition and restoration using total variation minimization and the H-1 norm,” Multiscale Model. Simul., vol. 1, no. 3, pp. 349–370, 2003.
\bibitem{b6} D. Scharstein and C. Pal, “Learning conditional random fields for
stereo,” in Proc. IEEE Conf. Comput. Vis. Pattern Recognit., Jun. 2007,
pp. 1–8.
\bibitem{b7} S. Isbaner, N. Karedla, D. Ruhlandt, S. C. Stein, A. Chizhik, I. Gregor,
and J. Enderlein, “Dead-time correction of fluorescence lifetime
measurements and fluorescence lifetime imaging,” Optics Express,
vol. 24, no. 9, pp. 9429–9445, May 2016, publisher: Optica Publishing
Group. [Online]. Available: https://opg.optica.org/oe/abstract.cfm?uri=
oe-24-9-9429.
\bibitem{b8} D. F. Yu and J. A. Fessler, “Mean and variance of single photon counting
with deadtime,” Physics in Medicine and Biology, vol. 45, no. 7, pp.
2043–2056, Jul. 2000.
\bibitem{b9}  J. Ma, S. Chan, and E. R. Fossum, “Review of Quanta Image
Sensors for Ultralow-Light Imaging,” IEEE Transactions on Electron
Devices, vol. 69, no. 6, pp. 2824–2839, Jun. 2022, conference
Name: IEEE Transactions on Electron Devices. [Online]. Available:
https://ieeexplore.ieee.org/document/9768129.
\bibitem{b10} M. -C. Amann, T. Bosch, M. Lescure, R. Myllyl¨a, and M. Rioux, “Laser
ranging: A critical review of usual techniques for distance measurement,” Opt. Eng., vol. 40, no. 1, pp. 10–19, 2001.

\end{thebibliography}
\end{document}